\documentclass[twocolumn]{aa}
\usepackage{graphicx}
\usepackage{txfonts}
\begin{document}
   \title{On the properties of contact binary stars}

   \subtitle{Research Note}

   \author{Sz. Csizmadia
          \inst{1}
          \and
          P. Klagyivik\inst{2}
          }

   \offprints{Sz. Csizmadia}

   \institute{Konkoly Observatory of the Hungarian Academy of Sciences, 
              H-1525 Budapest, P. O. Box 67., Hungary\\
              \email{csizmadia@konkoly.hu}
         \and
             Department of Astronomy of E\"otv\"os Lor\'and University,
             H-1112 Budapest, P\'azm\'any P. s\'et\'any 1/A, Hungary\\
             \email{klagyi@ludens.elte.hu}
             }

   \date{Received 12 March 2004; accepted 2 August 2004}

   \abstract{A catalogue of light curve solutions of contact binary stars 
             has been compiled. It contains the results of 159 light curve 
	     solutions. Properties of contact binary stars were studied 
	     by using the catalogue data. 

             As it is well known since Lucy's (1968a,b) and Mochnacki's (1981) 
	     works, primary components transfer their own energy to the 
	     secondary star via the common envelope around the two stars. This 
	     transfer was parameterized by a transfer parameter 
	     (ratio of the observed and intrinsic luminosities of the primary 
	     star). We proved that this transfer parameter is a simple
	     function of the mass and luminosity ratio. This 
	     newly found relation is valid for all systems except H type
	     systems which have a different relation.

             We introduced a new type of contact binary stars: H subtype 
	     systems which have a large mass ratio ($q>0.72$). These systems
	     show highly different behaviour on the luminosity ratio - 
             transfer parameter
	     diagram from other systems and according to our results the
	     energy transfer rate is less efficient in them than in other type
	     of contact binary stars. We also show that 
	     different types of contact binaries have well defined locations on the mass ratio 
	     - luminosity ratio diagram. All contact binary systems do not follow 
             Lucy's relation 
	     ($L_2/L_1 = (M_2/M_1)^{0.92}$). No strict mass ratio - 
             luminosity ratio 
	     relation of contact binary stars exists.

   }

   \maketitle
%
%________________________________________________________________

\section{Introduction}

Contact binary stars (or W UMa-type stars) consist of two dwarf stars whose 
spectral types are of F, G or K (only a few examples are known from earlier 
spectral types, and no M spectral type contact binary star is known). 
Binnendijk (1965) 
pointed out that the components in a contact binary system have nearly equal 
surface temperatures and luminosities in spite of their often highly
different masses. If 
they are really dwarf stars as suggested from their spectra, what is the 
mechanism which equalizes their temperatures and luminosities? The 
answer was 
given in Lucy's papers. The light curve characteristics of these binaries were 
succesfully interpreted by the contact model (Lucy 1968a,b), which 
simultaneously explains the shape of the light curve and the equal 
temperatures and luminosities of the components, hence this model is accepted
for describing theoretically W UMa-type stars. The model assumes that both stars fill 
their Roche-lobe and therefore they are touching each other. Mass and 
luminosity is transferred from the primary star to the secondary star through the narrow 
neck between the components. In the model two main sequence stars are 
embedded in a common photosphere which is convective. Although it explains the 
light curve shape and the temperature equalization, however, contact binary star
evolution and 
internal structure of the common photosphere remained open issues. Internal 
structure models were discussed by K\"ahler (1986) and a recent review on them can 
be found in Webbink (2003). K\"ahler (1989) summarized other possible theories of
contact binary stars and he found they were not approved by observational 
facts.
 Recent detailed computations on energy transfer
and internal structure were published by K\"ahler (2002a, 2002b, 2003).

In order to develop our empirical knowledge about these stars, a 
catalogue on the results of the light curve (LC) solutions of contact binary 
stars - based on LC-solutions published formerly in the literature - was 
compiled and it contains the solutions of light curve of 159 systems. This 
catalogue is presented in Sect. 2.

Based on the catalogue data, we investigated the 
efficiency of the energy transfer from the primary to the secondary star 
and the mass ratio - luminosity ratio relation. These are outlined in 
Sect. 3.

According to the contact model, the energy generated in the cores 
of the
components are redistributed in the common convective envelope and therefore 
the observable luminosities have another dependence on the mass ratio 
which would be the case if one looks at two main sequence stars in detached 
configuration (where $L_2 /L_1 = (M_2 / M_1)^{4.6}$). 
Lucy (1968a) found that the observable luminosity ratio is proportional to the 
ratio of the stellar surfaces ($L_2 /L_1 = (M_2 / M_1)^{0.92}$). We will show 
that the situation is more complex.

Mochnacki (1981) assumed that the energy transfer rate from the primary star to 
the secondary one depends only on the mass ratio. Liu \& Yang (2000) calculated 
the energy transfer rate and found its dependence on the mass ratio and on 
the evolutionary factor (which is defined 
as the ratio of the present radius and the zero-age radius of the primary
component). Recently K\"ahler (2002a,b) examined this question with detailed
numerical computations and he found the rate of the transferred luminosity
to be variable in time. The transferred luminosity can vary within wide 
limits (see Fig. 2 of K\"ahler 2002b) for a contact binary. 
(It should be emphasized that recent theoretical internal structure models of 
contact binary systems are inconsistent, as was described in K\"ahler 2002b.) 

Kalimeris \& Rovithis-Livaniou (2001) found that the observed rate of
energy transfer is a function of the secondary's luminosity. We also
examined empirically the energy transfer rate and found a simple relation
between the luminosity and mass ratio and the amount of the transferred
luminosity.

As the referee of this paper pointed out the width of the neck is
determined by the fill-out factor (which measures the degree of contact)
and W type systems have thinner neck than A-type ones (see e.g. Mochnacki
1981; of course, our sample confirms this establishment) and have higher
transfer rate and luminosity ratio than A-type systems (Figures~1 and 2).
One can mind that the thickness of the neck determines the amount of
transferred luminosity because through a thicker neck more luminosity can
be transported, but this does not realized. 

The reason of this paradoxon is not known yet. As a trend it is correct
that the thinner the neck, the larger the rate of transferred luminosity
which is exactly the opposite case than we expect. Note that K\"ahler's 
(2002b) numerical simulation showed that in general the degree of contact
varies in phase with the transferred luminosity (compare Figures~1 and 2 
in K\"ahler 2002b). The solution of this discrepancy between theory and 
observations requires further studies.

\section{The catalogue} 
%________________________________________________________________

There are two recent catalogues on contact binary stars (Maceroni \& van't
Veer, 1996; Pribulla et al., 2003). The catalogue of Maceroni \& van't
Veer (1996) lists 78 systems, while catalogue of Pribulla et al. (2003)
contains 361 field contact binaries. The contents of these and our
catalogues are demonstrated in Table~1 for comparison.

%__________________________________________________ One column table
   \begin{table}
      \caption[]{Comparison of contents of different catalogues of contact binaries. MV96: 
                 Maceroni \& van't Veer 1996, PKT03: Pribulla et al., 2003.}
         \label{KapSou1}
\small{
     $$   
\begin{array}{llll}
            \hline
            \noalign{\smallskip}
            \mathrm{Content}  &  \mathrm{MV96} & \mathrm{PKT03} & \mathrm{This~study} \\
            \noalign{\smallskip}
            \hline
\mathrm{photometric~mass~ratio}                                & X & X & X \\
\mathrm{spectroscopic~mass~ratio}                              & X & X & X \\
\mathrm{average~fractional~radii}                              & X &   & X \\
\mathrm{temperatures~of~the~components}                        & X & X & X \\
\mathrm{semiamplitude~of}~V_{rad}~\mathrm{of~the~primary~star} & X &   &   \\
\mathrm{magnitude~of~the~O'Connell~effect}                     & X & X &   \\
\mathrm{absolute~dimensions}~(R_{1,2}, M_{1,2}, L_{1,2}        & X &   &   \\
\mathrm{of~the~components)}                                    &   &   &   \\
\mathrm{system's~angular~momentum}                             & X &   &   \\
\mathrm{ephemeris~(epoch~and~period)}                          &   & X & X \\
\mathrm{number~of~available~minima~observations}               &   & X &   \\
\mathrm{information~about~period~change}                       &   & X &   \\
\mathrm{code~which~was~used~for~light~curve~solution}          &   & X &   \\
\mathrm{inclination}                                           &   & X & X \\
\mathrm{fillout~factor}                                       &   & X & X \\
\mathrm{spectral~type~and~distance}                            &   & X &   \\
\mathrm{range~of~variability}                                  &   & X &   \\
(m_1 + m_2) \sin^3 i                                           &   & X &   \\
\mathrm{fractional~luminosity~of~the~primary}                  &   &   & X \\
\mathrm{fractional~luminosity~of~the~third~body}               &   &   & X \\
\mathrm{dimensionless~surface~potentials}  &   &   & X \\
\mathrm{spot~parameters}                                       &   &   & X \\
\mathrm{gravity~darkening~exponents}                           &   &   & X \\
\mathrm{albedos~and~limb~darkening ~coefficients}              &   &   & X \\

            \noalign{\smallskip}
            \noalign{\smallskip}
            \hline
         \end{array}
     $$
}
   \end{table}

If recent ephemeris was not available we repeated the ephemeris given in
GCVS (Kholopov et al., 1998). Note that dimensionless surface potentials
are generally assumed to be equal for the two components - the exceptions
are noted in the catalogue. Gravity darkening exponents, albedos and
limb-darkening coefficients were generally fixed by the modellers with
some exceptions, and these exceptions are noted in the catalogue. The
catalogue together with its references is published in Table~2. 

For homogeneity, we primarily collected results of LC-solutions carried out 
by any version of the Wilson - Devinney code, however, in order to 
increase the sample results of modeling made by the BYNSIN Code (Vink\'o 
et al., 1992) was also included, but these are listed in a different table 
(see Table~3\footnote{Tables~2 and 3 are available only electronically at 
the homepage of the Konkoly Observatory via the URL: http://www.konkoly.hu}).

\section{Relation between astrophysical quantities} 
%________________________________________________________________

\subsection{Subtypes of contact binaries}

In 1965, contact binary stars were divided into two sub-types: A-type
systems (the larger star is the hotter one) and W-type systems (the
smaller star is the hotter one) (Binnendijk 1965). Later Lucy \& Wilson
(1979) introduced the terminus of B-type systems which are systems in
geometrical contact, but not in thermal contact and therefore there are
high surface temperature differences between the components. In this study
we call B-type systems which has $1000$K or larger surface temperature
difference between the components. Note that B-type systems are sometimes
mentioned as PTC (Poor Thermal Contact) systems (e.  g., in Rucinski \&
Duerbeck 1997). 

We do not use the terminus `E(arly)-type systems' which would mean the
contact binaries of O, B and A spectral types. In Figures 1-4 they
do not have any special position in contrast to H-type systems. This
conclusion is in agreement with K\"ahler's (1989) remark: from
observational viewpoint there is no difference between early and late
spectral type systems. 

In our sample we found 45 A, 13 B, 24 H and 77 W subtype systems. 

%----------------------------------------------------------- Mass r. - lum. r. 

   \begin{figure}
   \centering
   \resizebox{\hsize}{!}{\includegraphics[angle=-90]{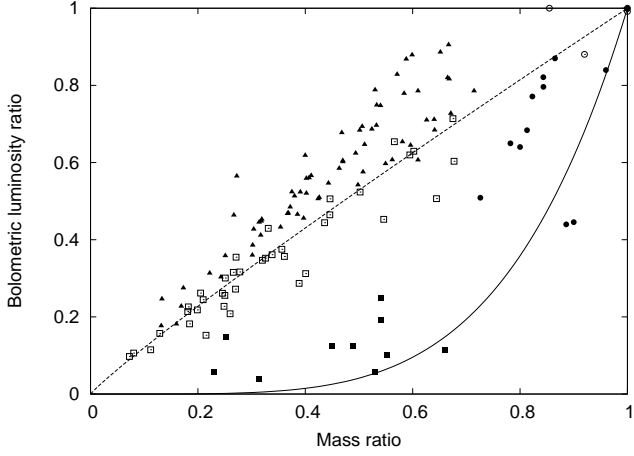}}
      \caption{Mass ratio - luminosity ratio diagram of contact binary
      stars. Open squares: A, filled squares: B, filled circles: H/A (A 
      subtype systems with $q>0.72$), open circles: H/W (W subtype 
      systems with $q>0.72$). filled triangles: W subtype systems. Solid line 
      is the $\lambda = q^{4.6}$, the main sequence mass-luminosity relation. 
      Dashed line  is Lucy's relation $\lambda=q^{0.92}$.}
         \label{figure1}
   \end{figure}

%----------------------------------------------------------- Beta. - lum. r. 

   \begin{figure}
   \centering
   \resizebox{\hsize}{!}{\includegraphics[angle=-90]{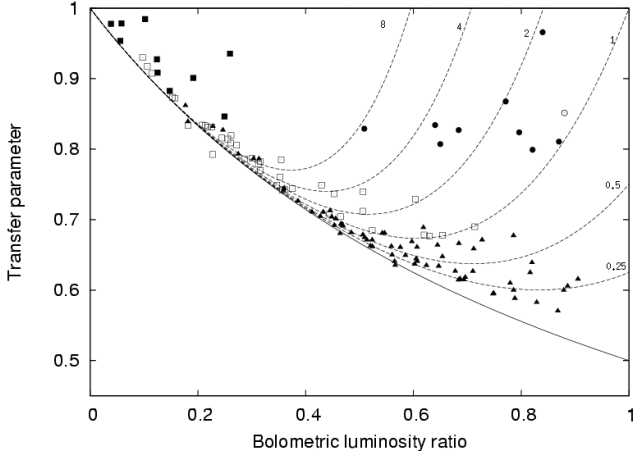}}
      \caption{Transfer parameter $\beta$ vs bolometric luminosity ratio 
      $\lambda$. Symbols are the same as in Figure~1. The solid line 
      describes the minimum rate of transfer parameter. Note 
      that most of the systems are on the this line. Systems which have   
      $q>0.72$ (H subtype systems) are far from this envelope. Dashed 
      lines are due to different values of $\alpha$ (see text for further 
      explanation) and the numbers show the corresponding values.} 
   \label{figure2}
   \end{figure}

%----------------------------------------------------------- Beta. - lum. r.2. 

   \begin{figure}
   \centering
   \resizebox{\hsize}{!}{\includegraphics[angle=-90]{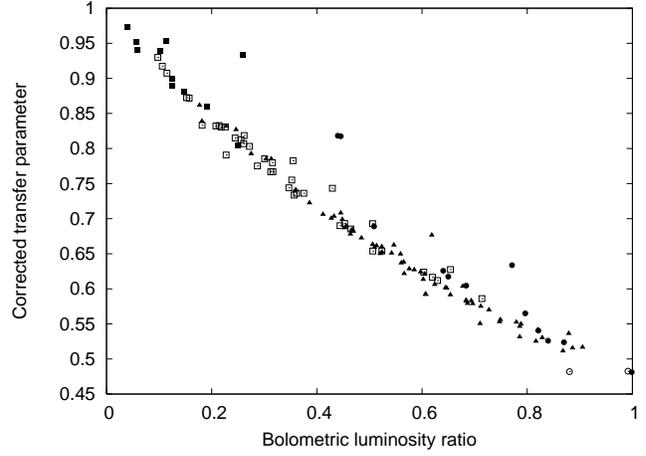}}
      \caption{Transfer parameter $\beta$ corrected with the mass ratio 
      vs bolometric 
      luminosity ratio $\lambda$. Symbols are the same as in Figure 1. 
      }
   \label{figure1}
   \end{figure}

   \begin{figure}
   \centering
   \resizebox{\hsize}{!}{\includegraphics[angle=-90]{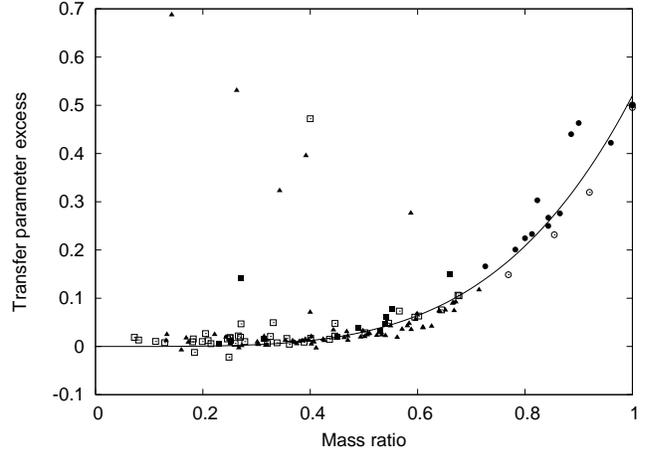}}
      \caption{Transfer parameter excess ($\beta - (1+\lambda)^{-1}$) vs 
               mass ratio. Symbols are the same as in Figure~1. Solid 
               line is the excess$ = 0.52 q^{4.1}$ fit.}
   \label{figure1}
   \end{figure}

\subsection{Mass ratio - luminosity ratio relation}

Several systems were excluded from the next analysis because some authors did not publish all
parameters of the light curve fitting. If fill-out factor was not given, we computed it with the
BinMaker 2.0 software (Bradstreet 1992) from the mass ratio and surface potential, but other
missing results could not be reproduced by us. That is why these systems were omitted from the
sample. 

The temperatures of the components were known from modelings, therefore we could calculate the
bolometric luminosity ratios from the measured ratios in $V$ band via $\lambda=(L_2/L_1)_{bol} =
(L_2/L_1)_{V} 10^{0.4\cdot(BC_1 - BC_2)}$. Bolometric luminosity ratio vs mass ratio
($q=M_2/M_1)$) is shown in Fig.~1. (Bolometric correction was calculated from Flower's (1996)
tables.)

The subtypes of W UMa-stars are located at different regions in this diagram. W subtype systems
have larger luminosity ratio than A subtype ones at a given mass ratio. This is natural because
the radius ratio is proportional to $q^{0.46}$ at a given mass ratio\footnote{This relation was
found by Kuiper (1941), and it is a natural consequence of the Roche-geometry assumption in light
curve models. We also checked this relation applying the catalogue data and $(R_2/R_1) = q^{0.459
\pm 0.003}$ was found.}, the temperature ratio $T_2 / T_1$ is larger in W systems and hence the
luminosity ratio is higher in W than in A systems. 

Both A and W subtype systems have larger luminosity ratios than any B subtype systems. A line
represents the mass-luminosity relation if the components were main-sequence stars (in this case
$\lambda = q^{4.6}$). Note that all systems are above this line (with three exceptions:
 W Crv and LP UMa, and the B-type system HW Per). B subtype systems are located
between A/W sytems and the $\lambda=q^{4.6}$ line. 

Another interesting fact is that A-type systems are relatively rare objects at high mass ratios,
but W-type systems show a completely opposite behaviour: they populate the region of higher mass
ratios, and there are only few W-type systems below $q=0.3$. This effect was discovered by
Maceroni et al. (1985) while studying properties of W UMa stars based on a sample containing 42
systems. We confirm their result using a much larger sample. 

\subsection{Energy transfer}

It is clear from Figure~1 that different mass ratio - luminosity ratio
relations exist in the case of different subtypes. Systems also show
remarkable diversity on the $\lambda$-$q$ plane. This is not due to the
scatter of data. The luminosity ratio can be determined with an accuracy
of 1-2\% or better, while precision of mass ratio measured photometrically
is generally better than a few per cent.  In general, spectroscopically
and photometrically determined mass ratios show a good agreement (Maceroni
\& van't Veer 1996, Pribulla et al. 2003). The diversity can be due to the
different rate of luminosity transferred from the primary to the secondary
star.  (K\"ahler, 2002b suggested such an effect: in the same contact
binary the transferred luminosity varies in time. At a given mass ratio
this can cause diversity in luminosity.)

We studied the energy transfer rate by the introduction of the 
transfer parameter. It was defined as 
\begin{equation}
\beta= \frac{L_{1,\mathrm{observed}}}{L_{1,\mathrm{ZAMS}}}
\end{equation}
It is easy to show that $\beta$ can be computed as 
\begin{equation}
\beta=\frac{1+q^{4.6}}{1+q^{0.92}(\frac{T_2}{T_1})^4}
\end{equation}
or
\begin{equation}
\beta=\frac{1+\alpha \lambda^{5.01}}{1+\lambda}
\end{equation}
where $\alpha = (T_1/T_2)^{20.01}$. Transfer parameter was plotted against
luminosity ratio (Fig.~2), and a good correlation was found with exception 
of all systems which have $q>0.72$ (they are marked by a 
different symbol in the Figures). The envelope is due to the minimum 
rate of the transfer parameter at a given luminosity ratio, so it is 
\begin{equation}
\beta_{\mathrm{envelope}}= \frac{1}{1+\lambda}
\end{equation}
It is interesting that systems with
$q>0.72$ are far from this envelope (hereafter we call these systems 
H-systems denoting high mass ratio systems) but other subtypes are close 
to it.

To quantify the deviation from Eq. (4) we calculated the transfer
parameter excess which was defined as the difference between $\beta$ 
and the envelope given by Eq. (4). The excess was found to be a function
of the mass ratio (Fig. 4) and a fit yielded that the he excess is 
proportional to $0.52 (\pm 0.02) q^{4.1} (\pm 0.1)$.

The systems should increase the luminosity transfer from the primary
to the secondary in order to equalize the surface temperatures if the mass
ratio is lower, but the situation is more complex than this simple
picture. Note that Kalimeris \& Rovithis-Livaniou (2001) found that the
transferred luminosity is the function of the secondary's luminosity. Our
results showed that the rate of the transferred luminosity is not related
only to the luminosity ratio but it is the function of the mass ratio,
too. 

In Figure~2 the $\beta - \lambda$ curves for different $\alpha$ values 
($\alpha = 0.25$, $0.5$, $1.0$, $2.0$, $4.0$ and $8.0$) are also shown. 
(Note that 
$\alpha$ depends strongly on the surface temperature ratio: $T_1 / T_2 = 
1.11$ corresponds to $\alpha \approx 8$.) The $\alpha$ values of A, B and W 
subtype systems generally are close to 1 while H subtype systems have 
larger temperature ratio. If the mass ratio is close to 1, we would wait 
that the two components have very similar features and hence their 
surface temperature ratio (and $\alpha$) is close to 1. In reality we 
found the opposite case: at large mass ratios the surface temperatures 
can be very different.

From the definition of the transfer parameter it is clear that the amount 
of the transferred luminosity is $\Delta L = (1-\beta) L_1$. Substituting 
$\beta = (1+\lambda)^{-1} - 0.52 q^{4.1}$ we found that $\Delta L = 
(\frac{\lambda}{1+\lambda} - 0.52q^{4.1}) L_1$.

Figure~3 shows the corrected $\beta$ ($\beta_{\mathrm{corr}} = \beta -
0.52q^{4.1}$) against bolometric luminosity ratio. The correlation 
between them is very good confirming our conclusion that transferred 
luminosity is a function of mass and luminosity ratios.

\section{Summary} 
%________________________________________________________________

The results of this research note can be summarized as follows.

\begin{itemize}

\item     [1] We compiled the catalogue of light curve solutions of contact
          binary stars. The catalogue contains LC-solutions of 159 systems.

\item     [2] We found that there is no strict mass ratio - luminosity ratio
          relation for contact binary stars. Such a relation was suggested by
	  Lucy (1968a), but K\"ahler's (2002b) results indicated large
	  luminosity ratio variations with small - practically unobservable by
	  light curve modeling - mass ratio variations. K\"ahler's model
	  does not contradict our results. 
\item	  [3] The
	  energy transfer from the primary star to the secondary star was found to
	  be depending on the mass ratio and the luminosity ratio. In H 
          systems the energy transfer rate is less
	  efficient than in other type ones at a given luminosity ratio.
          We determined the amount of the transferred luminosity from 
          the  primary to the secondary star and it was found a 
          function of the mass ratio and the luminosity ratio.

\end{itemize}
			 	 									
\begin{acknowledgements}									       
This work was supported by the Hungarian Science Fund under Grant Number T 034			       
551. This research has made use of the SIMBAD database, operated at CDS, 			       
Strasbourg, France; and of the NASA's Astrophysics Data System Abstract			       
and Article Service. We thank Drs L. Szabados and L. Patk\'os for their kind 
help and advice	during the preparation of the manuscript.							       
\end{acknowledgements}

\end{document}